\documentclass[apj]{emulateapj}
\usepackage{hyperref}
\usepackage{epsfig}
\usepackage[normalem]{ulem}
\usepackage[utf8]{inputenc}
\usepackage[T1]{fontenc}
\usepackage{graphicx}
\usepackage{captcont} 
\usepackage[english]{babel}
\usepackage{amssymb}
\usepackage{float}
\usepackage{multirow}
\usepackage{color}
\usepackage{threeparttable}
\usepackage{color}
\usepackage{arydshln}

\def\mearth{{\rm\,M_\oplus}}

\usepackage{times}
\slugcomment{}
 
\shorttitle{Survivor bias: ejected vs. surviving planetesimals}
\shortauthors{Raymond, Kaib, Armitage, \& Fortney}

\begin{document}

\title{Survivor bias: divergent fates of the Solar System's ejected vs. persisting planetesimals}


\author{Sean N. Raymond}
\affil{ Laboratoire d'Astrophysique de Bordeaux, Univ. Bordeaux, CNRS, B18N, allée Geoffroy Saint-Hilaire, 33615 Pessac, France; rayray.sean@gmail.com}

\author{Nathan A. Kaib}
\affil{HL Dodge Department of Physics Astronomy, University of Oklahoma, Norman, OK, USA}

\author{Philip J. Armitage}
\affil{Department of Physics and Astronomy, Stony Brook University, and Center for Computational Astrophysics, Flatiron Institute, both New York, USA}

\author{Jonathan J. Fortney}
\affil{Department of Astronomy and Astrophysics, University of California, Santa Cruz, USA}

\begin{abstract}
The orbital architecture of the Solar System is thought to have been sculpted by a dynamical instability among the giant planets. During the instability a primordial outer disk of planetesimals was destabilized and ended up on planet-crossing orbits. Most planetesimals were ejected into interstellar space but a fraction were trapped on stable orbits in the Kuiper belt and Oort cloud. We use a suite of N-body simulations to map out the diversity of planetesimals' dynamical pathways. We focus on two processes: tidal disruption from very close encounters with a giant planet, and loss of surface volatiles from repeated passages close to the Sun. We show that the rate of tidal disruption is more than a factor of two higher for ejected planetesimals than for surviving objects in the Kuiper belt or Oort cloud. Ejected planetesimals are preferentially disrupted by Jupiter and surviving ones by Neptune. Given that the gas giants contracted significantly as they cooled but the ice giants did not, taking into account the thermal evolution of the giant planets decreases the disruption rate of ejected planetesimals. The frequency of volatile loss and extinction is far higher for ejected planetesimals than for surviving ones and is not affected by the giant planets' contraction. Even if all interstellar objects were ejected from Solar System-like systems, our analysis suggests that their physical properties should be more diverse than those of Solar System small bodies as a result of their divergent dynamical histories. This is consistent with the characteristics of the two currently-known interstellar objects. 
\end{abstract}

\section{The Solar System's planetesimals}
The Solar System's small body populations represent the last planetesimals leftover from the planets' formation. They contain very little mass: the asteroids add up to less than a thousandth of an Earth mass~\citep{krasinsky02,kuchynka13}, the entire Kuiper belt perhaps a Mars mass~\citep{gladman01}, and the Oort cloud a few Earth masses at most~\citep{dones16}. Yet their initial mass budgets were likely significant larger, with up to a few $\mearth$ in the primordial asteroid belt~\citep[see discussion in][]{raymond18d} and $10-30 \mearth$ in the early Kuiper belt~\citep{nesvorny12}.
 
The Solar System's giant planets are thought to have undergone a dynamical instability~\citep{tsiganis05,morby07}. Current thinking suggests the following scenario. When the gaseous disk dissipated the giant planets were on more compact, resonant orbits, with an outer disk of planetesimals~\citep{morby07}. Interactions between the planets and planetesimal disk \citep[or perhaps simply among the planets; ][]{ribeiro20} triggered the instability~\citep{levison11,deienno17}, during which the Kuiper belt and Oort cloud were populated~\citep[][; note that the `cold classical' Kuiper belt is thought to have been left largely intact]{brasser13b,nesvorny17}. Originally proposed as a delayed event~\citep{gomes05}, a consensus is emerging that the instability happened early, no later than 100 Myr after the start of planet formation~\citep{zellner17,morby18,nesvorny18,mojzsis19,hartmann19}. 

The instability marks the emptying of large stable reservoirs of planetesimals via planetary scattering. The forces felt by planetesimals are not purely gravitational. Close encounters with giant planets -- especially very close ones -- can lead to tidal disruption~\citep[e.g.][]{asphaug96,richardson98}. Surface volatiles are lost during passages close to the Sun; some planetesimals lose their cometary activity and become extinct~\citep[e.g.][]{levison97,disisto09}, and sublimation-driven activity may also flatten comets' shapes~\citep{zhao20}.

Here we model the dynamical pathways of the planetesimals born in the primordial Kuiper belt. There is a significantly higher rate of tidal disruption and extinction among ejected planetesimals compared with surviving ones.  This may create physical differences between the size and surface distributions of surviving Solar System objects and interstellar objects, and also create a diversity of physical characteristics within each population. These processes have already been invoked to explain the properties of known interstellar objects~\citep{raymond18c,raymond18}.

\section{Dynamical simulations}

Our simulations were designed to capture the effect of the giant planets' instability, which we assume to have taken place shortly after the dispersal of the Sun's gaseous disk.  We were guided by previous studies of the instability that determined the initial conditions most likely to produce Solar System-like outcomes~\citep{nesvorny12}.  The four giant planets -- as well as a fifth ice giant that was included to increase the probability of Solar System-like outcomes~\citep{nesvorny11} -- were initially placed in a chain of mutual mean motion resonances (from the inside-out, in 3:2, 3:2, 2:1, and 3:2 resonance), anchored by Jupiter at 5.95 au and Neptune at 20.4 au.  An outer disk of 1000 planetesimals was placed on low-eccentricity, low-inclination orbits from 21.4 au (2 Hill radii exterior to Neptune's orbit) out to 30 au, constrained by planetesimal-driven migration studies to be the outer edge of the disk~\citep{gomes04}. The planetesimal disk contained a total of $20 \mearth$ and followed an $r^{-1}$ surface density profile. 

Each simulation was integrated for 1 Gyr using a version of the {\em Mercury} hybrid integrator~\citep{chambers99} that was modified to include the Galactic tidal field and perturbations from passing field stars~\citep{heisler86,rickman08,kaib18}. The code did not account for the dynamics of the Sun's birth cluster. The code recorded each time a planetesimal passed within 2.5 au of the Sun.  Particles were considered ejected when they reached 1 parsec (206,265 au) from the Sun. 

\begin{figure}
  \begin{center} 
  \leavevmode \epsfxsize=0.5\textwidth\epsfbox{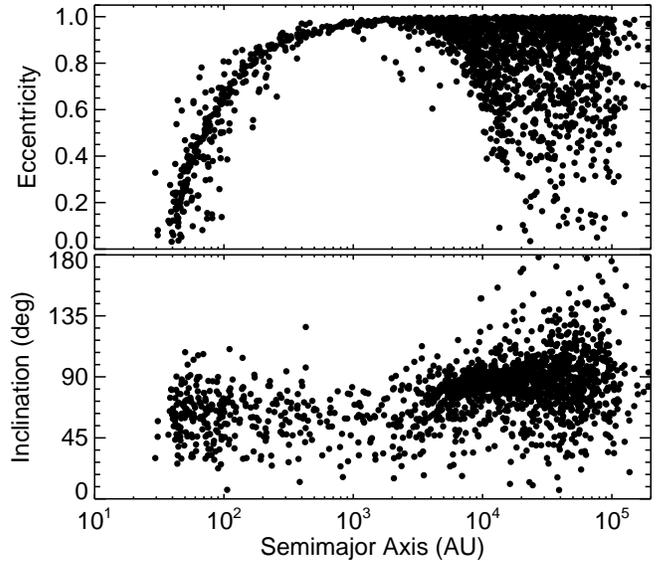}
    \caption[]{Orbital distribution of the surviving planetesimals in the simulations that provided a reasonable match to the giant planets. } 
     \label{fig:aei_all}
    \end{center}
\end{figure}

From our set of 100 simulations, 18 provided an acceptable match to the Solar System, with four surviving giant planets (in the right order) with orbital radii close to their current ones and eccentricities and inclinations within a factor of two. Our sample groups together particles from all 18 simulations.  A total of 15,104 planetesimals (83.9\%) were ejected, 634 (3.5\%) hit the Sun and 437 (2.4\%) collided with a planet.  The orbital distribution of the 1825 surviving planetesimals (10.1\%) in the 18 simulations is shown in Fig.~\ref{fig:aei_all}. The combined Oort cloud of these simulations contains 1518 particles for a cutoff of semimajor axes $a> 1000$~au, for an average Oort cloud mass of $1.68 \mearth$.  Surviving planetesimals within 1000 au are on orbits reminiscent of the Solar System scattered disk and contain an average of $\sim 0.3 \mearth$ within 1000 au (and $\sim 0.18 \mearth$ within 100 au). These populations are consistent with empirically-constrained estimates~\citep{gladman01,dones16}.

A first observation is that ejected planetesimals originated closer-in than planetesimals that remained on stable orbits around the Sun (that we refer to below as `survivors'). The most striking difference is between ejected planetesimals and scattered disk ($a<1000$~au) planetesimals. While the original planetesimal disk was less than 9 au in width, there was a 1.4 au difference in the median of these populations ($a_{init}$ of 25.6 au for ejected planetesimals and 27 au for scattered disk planetesimals), and a KS test found a probability of $p = 2 \times 10^{-8}$ that they were drawn from the same distribution.  The initial orbital radii of Oort cloud planetesimals were roughly consistent with those that were ejected ($p = 6 \times 10^{-2}$) and consistent with those that hit the Sun or a planet. This trend arises because, as we will see in the next Section, surviving planetesimals tend to avoid entering the inner Solar System and thus tend to originate farther away from the planets.  In contrast, the planetesimals that were ejected sample the outer disk uniformly.

\section{Tidal disruption}

The planetesimal disk is completely emptied by the giant planets' instability.  All simulated planetesimals underwent close encounters with at least one planet. The forces felt by planetesimals are not purely those of point mass gravity. Tidal disruption can occur when a planetesimal passes within a critical approach distance, which depends primarily on the planetesimal density~\citep[the rotation rate and tensile strength have second-order effects;][]{asphaug96,richardson98}. We adopt a simple formula for the tidal disruption radius $R_{tidal}$ from \cite{sridhar92}:
\begin{equation}
R_{tidal} = 1.69 \, R_{planet} \, \left(\frac{\rho_{planet}}{\rho_{planetesimal}}\right)^{1/3},
\end{equation}
\noindent where $R_{planet}$ is the planet's radius, and $\rho$ are bulk densities. 

The term {\em tidal disruption} encompasses a spectrum of outcomes. Numerical simulations have shown that a planetesimal that passes within $R_{tidal}$ will begin to shed mass but closer approaches are required for catastrophic disruption~\citep{richardson98,walsh18}. \cite{asphaug96} found that in a passage below $\sim 0.74 R_{tidal}$ [$0.55 R_{tidal}$] the largest surviving fragment will be smaller than 50\% [20\%] the size of the original body. We refer to the different disruption regimes as super-catastrophic ($d_{min}/R_{tidal} < 0.55$), catastrophic ($0.55<d_{min}/R_{tidal} < 0.74$) and ``gentle''  ($0.74< d_{min}/R_{tidal} < 1$). We expect outer disk planetesimals to have densities ${\rho \approx \rm 0.5 \ g \ cm^{-3}}$, the typical value measured or inferred from studies of specific comet nuclei~\citep[e.g.][]{asphaug94,carry12,patzold16}. Jupiter's tidal radius is $R_{tidal} = 2.34 R_{Jup}$. In 1992 comet Shoemaker-Levy 9 catastrophically disrupted after a closest approach of 1.33 $R_{Jup}$~\citep{sekanina94,movshovitz12}, corresponding to $0.57 R_{tidal}$.

\begin{figure}
  \begin{center} 
  \leavevmode \epsfxsize=0.5\textwidth\epsfbox{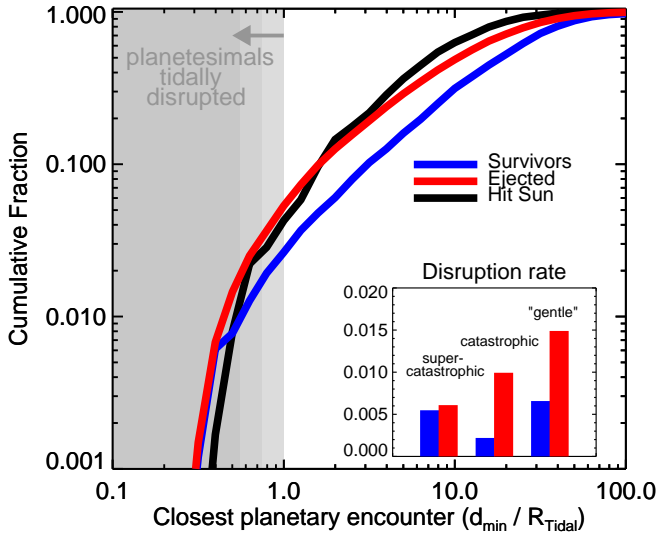}
    \caption[]{The closest encounters between planetesimals and planets.  The main plot shows the cumulative distribution of planetesimals' closest encounters with a planet $d_{min}$ normalized to the tidal disruption radius $R_{tidal}$, which was calculated assuming $\rho = {\rm 0.5 g \ cm^{-3}}$. The inset shows the rate of different regimes of disruption among ejected and surviving planetesimals: super-catastrophic ($d_{min}/R_{tidal} < 0.55$), catastrophic ($0.55<d_{min}/R_{tidal} < 0.74$) and ``gentle''  ($0.74< d_{min}/R_{tidal} < 1$).
    }
     \label{fig:minrt}
    \end{center}
\end{figure}

Figure~\ref{fig:minrt} shows the distribution of the closest approach that each planetesimal underwent throughout its evolution, $d_{min}$, normalized to $R_{tidal}$ of that planet. The overall tidal disruption rate is more than a factor of two higher among ejected planetesimals than surviving ones (3.1\% vs. 1.4\%). This is because ejected planetesimals were preferentially scattered by Jupiter and surviving ones by Neptune. The number of close encounters with Jupiter per tidal disruption event was 2640. For Neptune this value was 9 times higher. The reason for this difference is both geometrical and dynamic.  A close encounter takes place when a planetesimal enters a planet's Hill sphere ($R_{H} = a \, (M/3 M_\star)^{1/3}$, where $a$ is the orbital distance and $M$ the planet mass). The `disruption' part of parameter space can be thought of as a ring of trajectories between the planet's surface and $R_{tidal}$, and the probability of disruption scales with the surface area of that ring relative to that of the Hill sphere. By this fact alone, planetesimals are 32 times more likely to disrupt during an encounter with Jupiter than one with Neptune, although this factor drops to 17 by accounting for the fact that Neptune was closer to the Sun when most disruptions took place, at 20-24 au depending on the simulation.  Yet planetesimals encountered Neptune at much lower speeds than Jupiter. One would expect encounter velocities to scale with the orbital speed, yet in our simulations planetesimal encounter velocities were significantly faster compared with Neptune, roughly a factor of two (on average) higher than the ratio of the planets' orbital speeds. This is because by the time planetesimals encountered Jupiter they had already undergone a series of scattering events with other planets that acted to increase their random velocities. Gravitational focusing -- which scales as $1 + (v_{esc}/v_{rand})^2$, where $v_{esc}$ is the planet's escape speed and $v_{rand}$ a planetesimal's random velocity -- was almost twice as strong for Neptune than for Jupiter. This combination of geometry and gravitational focusing predicts a factor of $\sim 9$ times higher disruption rate for Jupiter than for Neptune, in agreement with the simulations.

The rates of catastrophic and super-catastrophic disruption were closer for ejected and surviving planetesimals (inset in Fig.~\ref{fig:minrt}).  Yet given small number statistics, with just 26 total disruption events among 1825 surviving planetesimals, exploring the relative distribution of disruption events is left for future study.

\section{The role of giant planet contraction}

The giant planets contracted and cooled after their formation~\citep[e.g.][]{marley07,fortney11}. One may wonder whether many tidal disruption events inferred from planet-planetesimal close encounters should really have resulted in collisions. To test this, we adopted radius evolution models from \cite{fortney11} for all four giant planets. (The disruption rate of planetesimals from encounters with the third, ejected ice giant was low enough to ignore). These models calculate the radiative atmosphere and convective interior cooling and planet-wide contraction of each Solar System giant planet.  While they are constrained to match the present-day planets, they do not contain a formation model.  Adiabatically ``rewinding the clock'' to just after their formation likely overestimates their early sizes by 10-20\%~\citep{marley07,fortney11}, so including these models likely modestly overestimates the effect of contraction.

\begin{figure}[t]
  \begin{center} 
  \leavevmode \epsfxsize=0.5\textwidth\epsfbox{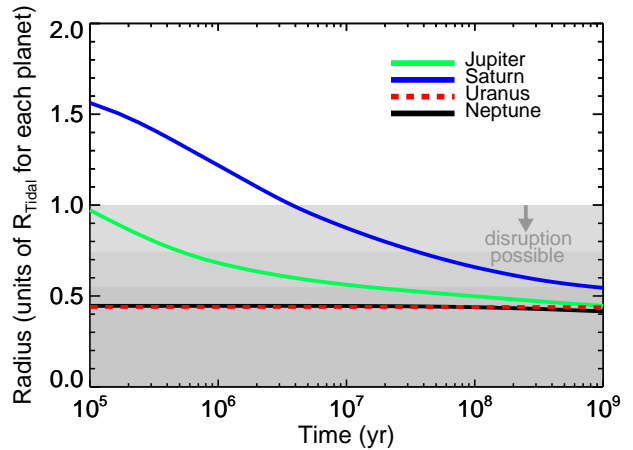}
    \caption[]{Model evolution of the radius of each giant planet~\citep[from][]{fortney11} in units of its tidal radius $R_{tidal}$. Tidal disruption is possible when a planet's physical radius becomes smaller than its tidal disruption radius (shading indicates different degrees of disruption).} 
     \label{fig:contraction}
    \end{center}
\end{figure}

Figure~\ref{fig:contraction} shows the evolution of model radii in units of each planet's tidal disruption radius $R_{tidal}$, which depends solely on the planet's mass and so remains fixed in time (Eq. 1).  Jupiter contracts on a $\sim 1$~Myr timescale and Saturn on $\sim 10$~Myr timescale, but the ice giants' radii barely change. Accounting for the giant planets' contraction therefore affects the disruption of ejected planetesimals but not survivors'. This is because, while ejected planetesimals underwent many close encounters with Jupiter and Saturn, surviving planetesimals tended to avoid the gas giants (Sec. 3). Tidal disruption events can only occur when a planet's radius drops below $R_{tidal}$, such that early disruptive encounters are replaced by planetary collisions. The rate of tidal disruption of ejected planetesimals dropped by $\sim$~20\%, as 58 tidal disruption events with Jupiter and 26 with Saturn should in principle have resulted in collisions. An even higher fraction of very close encounters -- those that lead to catastrophic disruption -- should have been collisions. Roughly 30\% and 43\% of encounters meeting the criteria for catastrophic and super-catastrophic disruption~\citep[$d_{min}/R_{tidal} < 0.74$ and $0.55$;][]{asphaug96}, respectively, should have resulted in collisions. Disruption events among surviving planetesimals were almost entirely from encounters with Neptune and none was affected by its contraction. In fact, after accounting for the giant planets' contraction, there were slightly more super-catastrophic disruption events among surviving planetesimals, although we caution that model uncertainties and small number statistics make this uncertain.

\section{Volatile loss and planetesimal extinction}

After repeated passages close to the Sun comets lose their surface volatiles and go extinct. They no longer outgas as they approach the Sun and become much fainter and more difficult to detect~\citep[see][for current constraints on the size distribution of cometary nuclei]{boe19}. Nonetheless, many extinct comet nuclei have been discovered~\citep[including the Damocloid population; see ][]{jewitt05}. While the timescales are too long to observe extinction in individual comets, dynamical models of the population of comets constrain the conditions that lead to extinction~\citep{levison97,disisto09,rickman17}. \cite{nesvorny17} found that the distributions of Halley-type and ecliptic comets was matched if extinction occurred after $N$ orbits with perihelion distance $q < 2.5$~au, where $N \sim 500-1000$.  

\begin{figure}[t]
  \begin{center} 
  \leavevmode \epsfxsize=0.5\textwidth\epsfbox{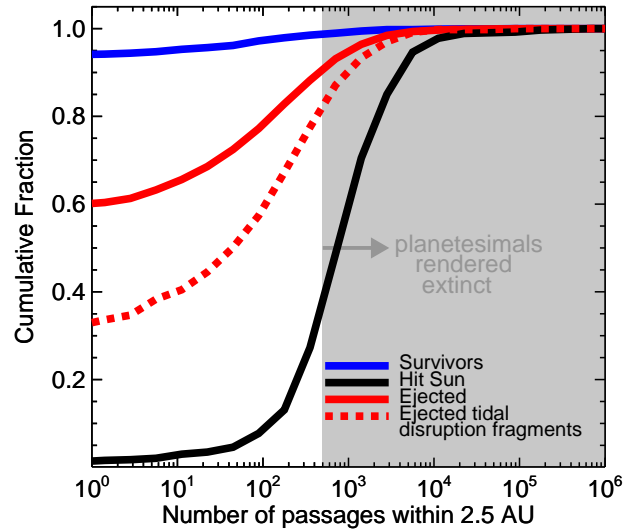}
    \caption[]{Distribution of the number of times a planetesimal passed within 2.5 au. The dashed red line shows the number of passages within 2.5 au for disrupted planetesimals, after disruption and before ejection. Curves do not start at the origin because many planetesimals never passed within 2.5 au.} 
     \label{fig:n25}
    \end{center}
\end{figure}

Figure~\ref{fig:n25} shows the distribution of the number of passages within 2.5~au of the Sun for different populations of planetesimals. Only 2.1\% of surviving planetesimals underwent 500 or more passages within 2.5 au and would have become extinct according to the criterion of \cite{nesvorny17}. In contrast, 17.1\% of ejected planetesimals would have been de-volatilized. The sub-population that spent the least time in the inner Solar System were surviving planetesimals that survived on orbits interior to 1000 au, for which only 1.6\% went extinct and only 3.6\% ever entered within 2.5 au of the Sun at all during the billion-year integrations. The process of ejection typically requires tens of close encounters with Jupiter. While ejected planetesimals are those with a net gain in orbital energy, the encounter geometry is random and some close encounters can kick planetesimals inward. Jupiter-family comets are representative of this behavior, as they were scattered inward by Jupiter but are on their way to ejection.  Extinct planetesimals are simply those that had a prolonged stay in the inner Solar System before being ejected. The fraction of planetesimals that was rendered extinct was barely affected when the giant planets' contraction was taken into account (as in Sec 4.).  Planetesimals that were rendered extinct passed closer to Jupiter than the average ejected planetesimal, with a closest encounter of $10 R_{tidal}$ for extinct ejected planetesimals vs. $69 R_{tidal}$ for all ejected planetesimals. Yet only a very small fraction (less than 2\%)  passed close enough to the giant planets' surfaces for their contraction to make a difference.

To match the ratio of new to returning long-period comets, \cite{nesvorny17} found that Oort cloud comets could not undergo more than 10 passages within 2.5 au without becoming extinct.  They attributed the more rapid extinction of these comets to their nuclei being much smaller than those of typical ecliptic or Halley-type comets~\citep[also proposed by][]{brasser13b}.  

Our simulated planetesimals are far more massive than real ones such that one particle represents a size distribution. After a planetesimal tidally disrupts into fragments its effective size distribution is shifted to much smaller bodies~\citep[see][and discussion below]{raymond18c}. Following \cite{nesvorny17} we impose an extinction criterion of just 10 passages within 2.5 au after a planetesimal has disrupted. The majority (65\%) of disrupted ejected planetesimals meet this criterion such that their fragments should have lost their volatiles prior to ejection and be extinct. A larger fraction of disrupted ejected planetesimals were affected by the giant planets' contraction, given that these are the planetesimals that underwent very close approaches to the giant planets.  Yet even after taking contraction into account, more than half (58\%) of fragments were de-volatilized prior to ejection.  In contrast, only 14\% of surviving disrupted planetesimals (just 4 particles) were rendered extinct, even with this much more lenient criterion. The reason is simply that the surviving planetesimals that were disrupted were mainly disrupted by Neptune and never entered the inner Solar System.  Indeed, three of the four disrupted surviving planetesimals that were rendered extinct had been disrupted by Jupiter (and the fourth by Uranus). 

Some dynamically new comets start outgassing when they pass within $\sim 30$~au of the Sun~\citep{meech09,sarneczky16}.  Extremely volatile species such as CO may drive this outgassing, and these ``supervolatiles'' may thus be lost after a relatively small number of passages through the inner Solar System. We can use our simulations to roughly evaluate the retention of supervolatiles among different populations of planetesimals (although it would be preferable to directly couple a thermal evolution code with planetesimals' orbital evolution; Gkotsinas in prep.).  We simply assume that a single encounter with Saturn indicates that a planetesimal spent enough time within $\sim 10$~au to lose all of its supervolatiles.  While this is a massive oversimplification~\citep[e.g., see][]{guilbert12}, we can already see clear trends.

Surviving planetesimals are far more likely to hold on to their supervolatiles than ejected ones.  The vast majority (87\%) of ejected planetesimals underwent at least one close encounter with Saturn, and most (61\%) also underwent encounters with Jupiter.  In contrast, only 37\% of surviving planetesimals ever encountered Saturn and only 10\% encountered Jupiter.  There are also significant differences among surviving planetesimals. Survivors within 1000~au were much more likely to retain supervolatiles than those in the Oort cloud. They encountered the gas giants at a much lower rate: 22\% [4.9\%] of planetesimals within 1000~au encountered Saturn [Jupiter], compared with 40\% [11.8\%] beyond 1000~au.

One might wonder: where in the Solar System can we find the most pristine, least altered planetesimals?  The cold classical Kuiper belt is an obvious answer, as these objects' dynamically cold orbits indicate that they may never have been scattered by planets~\citep[see][]{nesvorny15}.  Our simulated planetesimals that underwent no more than 100 planetary encounters and never entered the inner Solar System were preferentially found in two distinct areas: in the scattered disk just past Neptune, with orbital semimajor axes $a < 100$~au, and in the heart of the Oort cloud, with $a = 10^{4-5}$~au. 

\section{Discussion}

While they originated from the same parent population, planetesimals that were ejected from the Solar System had a different dynamical experience than surviving ones. We call this {\em survivor bias}. 

Compared with surviving planetesimals, a higher fraction of ejected ones underwent very close encounters with a giant planet. At face value this indicates a higher rate of tidal disruption among ejected planetesimals. However, the difference in disruption rates shrinks -- and may even reverse -- when the giant planets' contraction is taken into account. This is because ejected planetesimals' closest encounters were usually with Jupiter and surviving ones with Neptune, coupled with the fact that Jupiter's contraction is far more significant than Neptune's~\citep[][see Fig.~\ref{fig:contraction}]{fortney11}. For tidal disruption to play an important role in the size distribution, disrupted planetesimals must on average produce tens to hundreds of fragments. Comet Shoemaker-Levy 9 disrupted into $\sim 20$ visible fragments~\citep{scotti93,sekanina94}. Yet several families of Sun-grazing comets contains hundreds of members~\citep[e.g., the Kreutz, Marsden and Kracht groups -- see][]{knight10,lamy13}, each associated with a single parent body and presumably produced by tidal disruption. If tidal disruption fragments dominate a planetesimal population -- either among ejected or surviving bodies -- this should result in a steeper (i.e., more bottom-heavy) size distribution~\citep[for discussion of the expected size distribution of interstellar objects, see][]{rafikov18,raymond18c,moromartin18}.

Survivor bias implies that Solar System small bodies may not always provide good analogs for interstellar objects~\citep[assuming those to have a Solar System-like origin; see][]{issi19}. In addition, even coming from a unique parent population, not all interstellar objects should look alike. Based on simple criteria related to the number of passages within 2.5 au, a large fraction of planetesimals (17\%) and fragments (65\%) should have lost their volatiles on the pathway to ejection~\citep[see also][]{raymond18}. While we did not model it, the increased outgassing from passages close to the star may also change planetesimals' and fragments' shapes and effectively stretch them out~\citep{seligman20,zhao20}. A larger fraction of ejected planetesimals should also have lost their supervolatiles. These extinct objects might appear similar to `Oumuamua~\citep[indeed,][proposed that `Oumuamua represents an extinct cometesimal fragment]{raymond18c}, given its photometric similarity to volatile-rich Solar System objects but lack of visible activity~\citep{meech17,fitzsimmons18}. However, this model struggles to explain `Oumuamua's non-gravitational acceleration~\citep[see][]{seligman20}. In contrast, Borisov could simply represent a planetesimal that was ejected from its home system a bit more quickly, without repeated passages close to its star.  The most pristine planetesimals in the Solar System are likely to be trapped in the scattered disk or in the heart of the Oort cloud.

The giant planets were bombarded during the instability, enriching their atmospheres with solids~\citep{matter09}.  In our simulations, an average of 0.20--0.08--0.06--0.14~$\mearth$ in planetesimals collided with Jupiter--Saturn--Uranus--Neptune. When taking the planets' contraction into account, an additional $0.25 \mearth$ collided with Jupiter and $0.1 \mearth$ with Saturn.

\section*{Acknowledgments}
We thank the referee Darryl Seligman for a thorough report that greatly improved the paper.
SNR thanks the CNRS's PNP program for funding and fellow members of the ISSI 'Oumuamua team for helpful discussions. NAK acknowledges support from NASA Emerging Worlds Grant 80NSSC18K0600, NSF award AST-1615975, and NSF CAREER award 1846388. The computing for this project was performed at the OU Supercomputing Center for Education \& Research (OSCER) at the University of Oklahoma (OU).


\end{document}